\documentclass[10pt,aps,prl,twocolumn,showpacs,amsmath,amsmath,amssymb,superscriptaddress]{revtex4-1}
\usepackage{amssymb}
\usepackage{graphicx}
\usepackage{dcolumn}
\usepackage{bm}
\usepackage{psfrag}
\usepackage{bbold}
\usepackage{float}
\usepackage{epstopdf}

\begin{document}

\title{Collectively enhanced chiral photon emission from an atomic array near a nanofiber}

\author{Ryan Jones}
\affiliation{School of Physics and Astronomy and Centre for the Mathematics and Theoretical Physics of Quantum Non-Equilibrium Systems, The University of Nottingham, Nottingham, NG7 2RD, United Kingdom}

\author{Giuseppe Buonaiuto}
\affiliation{School of Physics and Astronomy and Centre for the Mathematics and Theoretical Physics of Quantum Non-Equilibrium Systems, The University of Nottingham, Nottingham, NG7 2RD, United Kingdom}

\author{Ben Lang}
\affiliation{School of Physics and Astronomy and Centre for the Mathematics and Theoretical Physics of Quantum Non-Equilibrium Systems, The University of Nottingham, Nottingham, NG7 2RD, United Kingdom}

\author{Igor Lesanovsky}
\affiliation{School of Physics and Astronomy and Centre for the Mathematics and Theoretical Physics of Quantum Non-Equilibrium Systems, The University of Nottingham, Nottingham, NG7 2RD, United Kingdom}
\affiliation{Institut f\"ur Theoretische Physik, Universit\"at Tübingen, Auf der Morgenstelle 14, 72076 T\"ubingen, Germany}

\author{Beatriz Olmos}
\affiliation{School of Physics and Astronomy and Centre for the Mathematics and Theoretical Physics of Quantum Non-Equilibrium Systems, The University of Nottingham, Nottingham, NG7 2RD, United Kingdom}
\affiliation{Institut f\"ur Theoretische Physik, Universit\"at Tübingen, Auf der Morgenstelle 14, 72076 T\"ubingen, Germany}

\date{\today}

\begin{abstract}
Emitter ensembles interact collectively with the radiation field. In the case of a one-dimensional array of atoms near a nanofiber, this collective light-matter interaction does not only lead to an increased photon coupling to the guided modes within the fiber, but also to a drastic enhancement of the chirality in the photon emission. We show that near-perfect chirality is already achieved for moderately-sized ensembles, containing 10 to 15 atoms. This is of importance for developing an efficient interface between atoms and waveguide structures with unidirectional coupling, with applications in quantum computing and communication such as the development of non-reciprocal photon devices or quantum information transfer channels.
\end{abstract}

\pacs{}

\maketitle

\textit{Introduction.} The radiative properties of a group of emitters are determined by the electromagnetic field mode structure of their enviroment \cite{dicke1954,agarwal1970,lehmberg1970,keaveney2012,pellegrino2014,bettles2016,sutherland2016,araujo2016,guerin2016}. They can be modified, for example, by the presence of nearby metallic or dielectric surfaces and nanospheres, metamaterials or plasmonic nanowires, among others \cite{dung2002,wallquist2009,zhou2011,martincano2011,stehle2014,hou2014,chao2016,jones2018}. This phenomenon, first described by Purcell in the 1940s \cite{purcell1946}, has been studied extensively in a variety of contexts, and most prominently in systems involving quantum optical devices \cite{spreeuw2005,fort2008,asenjo2017,kockum2018}.

Structured environments such as photonic crystals and optical fibers support a finite number of \emph{guided} electromagnetic field modes. These are particularly interesting as they can possess a significant longitudinal field component \cite{yariv1997,balykin2004,lekien2004,chen2010,bliokh2015}. This leads to the field around the fiber having elliptical polarization, whose sign depends on the direction of propagation of the mode. If the polarization of a nearby emitter is aligned with that of a guided mode, the emission will occur predominantly into this mode, travelling in either the forwards or backwards direction along the fiber. This so-called \emph{chiral} coupling has been observed experimentally for circularly polarized atoms near an optical fiber of sub-wavelength thickness (herein referred to as a nanofiber) \cite{mitsch2014,scheucher2016,solano2017,dareau2018}, as well as for a variety of other emitter types coupled with guided structures \cite{mrowinski2019, scarpelli2019}.

While it is well understood how the radiative properties of a single atom are altered  by the presence of a nanofiber \cite{klimov2004,lekien2005,lekien2014,coles2016}, much less is known about the behavior of atomic ensembles \cite{dzsotjan2010,dzsotjan2011,kornovan2016,lekien2017,mirza2017}. However, understanding this situation is of increasing importance, as collections of emitters near a nanofiber promise applications, e.g., in quantum information routing and processing \cite{lekien2008,stannigel2011,stannigel2012,lodahl2015,ramos2014,pichler2015,vermersch2017,jen2019,buonaiuto2019}. Moreover, the collective dissipative dynamics resulting from a competition between the coupling to the unguided modes of the free space and the guided ones of the nanofiber may result in the formation of complex many-body phases and phase transitions \cite{asenjo2017,buonaiuto2019}. 

In this paper, we explore the question of whether chirality can be enhanced due to collective effects. To this end, we investigate the photon emission from a weakly driven one-dimensional array of atoms in the vicinity of a nanofiber. We show that even for moderate number of atoms the majority of photons is emitted into the fiber with near perfect chirality. This enhanced coupling is mediated by the appearance of a collective superradiant mode, which forms due to the presence of the nanofiber. When the laser driving field matches the phase profile of this mode, a dramatic increase in the efficiency of the atom-fiber coupling and at the same time complete unidirectionality of the photon emission is achieved. These results are of immediate relevance to current experimental efforts aiming to control light-matter interactions through the use of nanophotonic structures, with applications in quantum information and communication.

\begin{figure}
\includegraphics[width=\columnwidth]{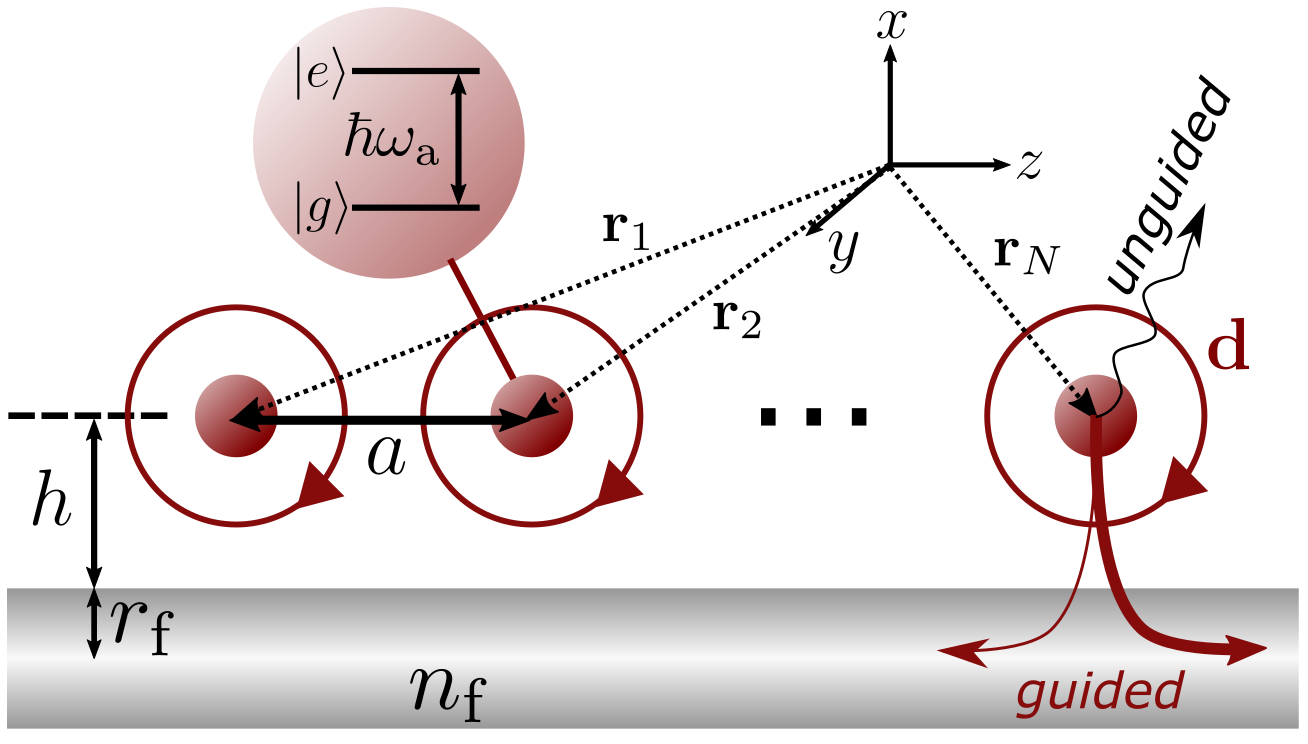}
\caption{\textbf{System.} A chain of $N$ identical two-level atoms (transition energy $\hbar \omega_\mathrm{a}$) with nearest neighbors separation $a$, is placed at a distance $h$ above a nanofiber, which has radius $r_\mathrm{f}$ and refractive index $n_\mathrm{f}$. The dipole moment of the transition $\left|g\right>\to\left|e\right>$ in each atom is $\mathbf{d}= (i,0,-1) d / \sqrt{2}$. The atoms are coupled both to the free field (unguided) and the nanofiber (guided) modes.}\label{fig:System}
\end{figure}

\textit{System and master equation.} We consider a chain of $N$ identical atoms aligned parallel to the $z$-axis, with lattice constant $a$ (see Fig. \ref{fig:System}). The internal structure of each atom is modelled as a two-level system, with ground $\left|g\right>$ and excited $\left|e\right>$ states separated by an energy $\hbar \omega_\mathrm{a}=\hbar 2\pi c/\lambda_\mathrm{a}$, where $\lambda_\mathrm{a}$ is the wavelength of the $\left|g\right>\to\left|e\right>$ transition. The atoms are placed at a distance $h$ above the surface of a cylindrical nanofiber with radius $r_\mathrm{f}$ and refractive index $n_\mathrm{f}>1$. In cylindrical coordinates $(r,\phi,z)$ the position of the atoms is given by $\mathbf{r}_{j} = (r_\mathrm{f} + h, 0, (j - 1)a)$ for $j = 1,2,...,N$. The dipole moments $\mathbf{d}$ of the transition $\left|g\right>\to\left|e\right>$ are considered identical for all atoms.

By solving Maxwell's equations, it can be shown that a nanofiber supports a small number of guided modes. We focus in the regime where the radius of the nanofiber obeys the so-called \emph{single-mode condition} \cite{snyder1983} $r_\mathrm{f}<2.405 \lambda_\mathrm{a} / (2 \pi \sqrt{n_\mathrm{f}^2 - 1})$, such that the only modes supported by the nanofiber are the hybrid fundamental $\mathrm{HE}_{11}$. Throughout, we assume the nanofiber is made from silica, and we calculate the refractive index $n_\mathrm{f}$ as a function of the atomic transition wavelength $\lambda_\mathrm{a}$ using the Sellmeier equation \cite{malitson1965}.

Under the Born and Markov approximations, one can obtain a quantum master equation, $\dot{\rho} = -\frac{i}{\hbar} \left[ \mathcal{H}, \rho \right] + \mathcal{D}(\rho)$, that describes the dynamics of the atoms through the reduced density matrix $\rho$ (see, e.g., \cite{lekien2017}). The first term on the right hand side of this equation describes coherent dipole-dipole interactions among the atoms that arise from the exchange of virtual photons. The Hamiltonian that determines this coherent dynamics is
\begin{equation}\label{eq:H}
\mathcal{H} = - \hbar \sum^N_{i \neq j=1} V_{ij} \sigma^\dag_i \sigma_j,
\end{equation}
where $\sigma_i = | g \rangle_i \langle e |$ denotes the lowering operator for the $i$-th atom. The rate of the dipole-dipole exchange between a pair of atoms $i$ and $j$ is characterised by the coefficient $V_{ij}$. The second term of the master equation encapsulates the action of dissipation in the system, and has the form
\begin{equation} \label{eq:Dissipator}
\mathcal{D}(\rho) = \sum_{ij=1}^N \Gamma_{ij}\left(\sigma_{j} \rho \sigma^\dag_i - \frac{1}{2} \{ \sigma^\dag_i \sigma_j, \rho \}\right).
\end{equation}
For a non-interacting chain of atoms, $\Gamma_{ij} = 0$ for $i \neq j$ such that each atom decays independently with rate $\Gamma_{ii}$ which, due to the presence of the nanofiber, can vary significantly from the decay rate in vacuum, $\gamma$. However, in general, $\Gamma_{ij} \neq 0$ for $i \neq j$ (e.g. in a dense atomic chain in free space \cite{bettles2016,sutherland2016} or near a nanofiber \cite{asenjo2017,lekien2017}), and the emission of photons from the chain becomes a collective process. This can be better understood by diagonalizing the coefficient matrix $\Gamma_{ij} = \sum_c M^\dag_{i c} \gamma_c M_{c j}$. The dissipator \eqref{eq:Dissipator} then assumes the diagonal form
\begin{equation}\label{eq:DissipatorDiag}
\mathcal{D}(\rho) = \sum_{c=1}^N \gamma_c \left( J_c \rho J^\dag_c - \frac{1}{2} \{ J^\dag_c J_c, \rho \} \right).
\end{equation}
Here, it is apparent that the emission occurs via the collective jump operators $J_c = \sum_j M_{c j} \sigma_j$, which in general consist of superpositions of all single-atom lowering operators. The collective decay rates $\gamma_c$ (the eigenvalues of the matrix of $\Gamma_{ij}$ coefficients) can be much larger or smaller than $\gamma$, which is commonly referred to as superradiant and subradiant decay, respectively \cite{dicke1954}. The exact form of $V_{ij}$ and $\Gamma_{ij}$, given in Appendix A, depend strongly on the system parameters, such as $a$, $\lambda_\mathrm{a}$, $r_\mathrm{f}$ and $h$.

\begin{figure}[ht]
\includegraphics[width=\columnwidth]{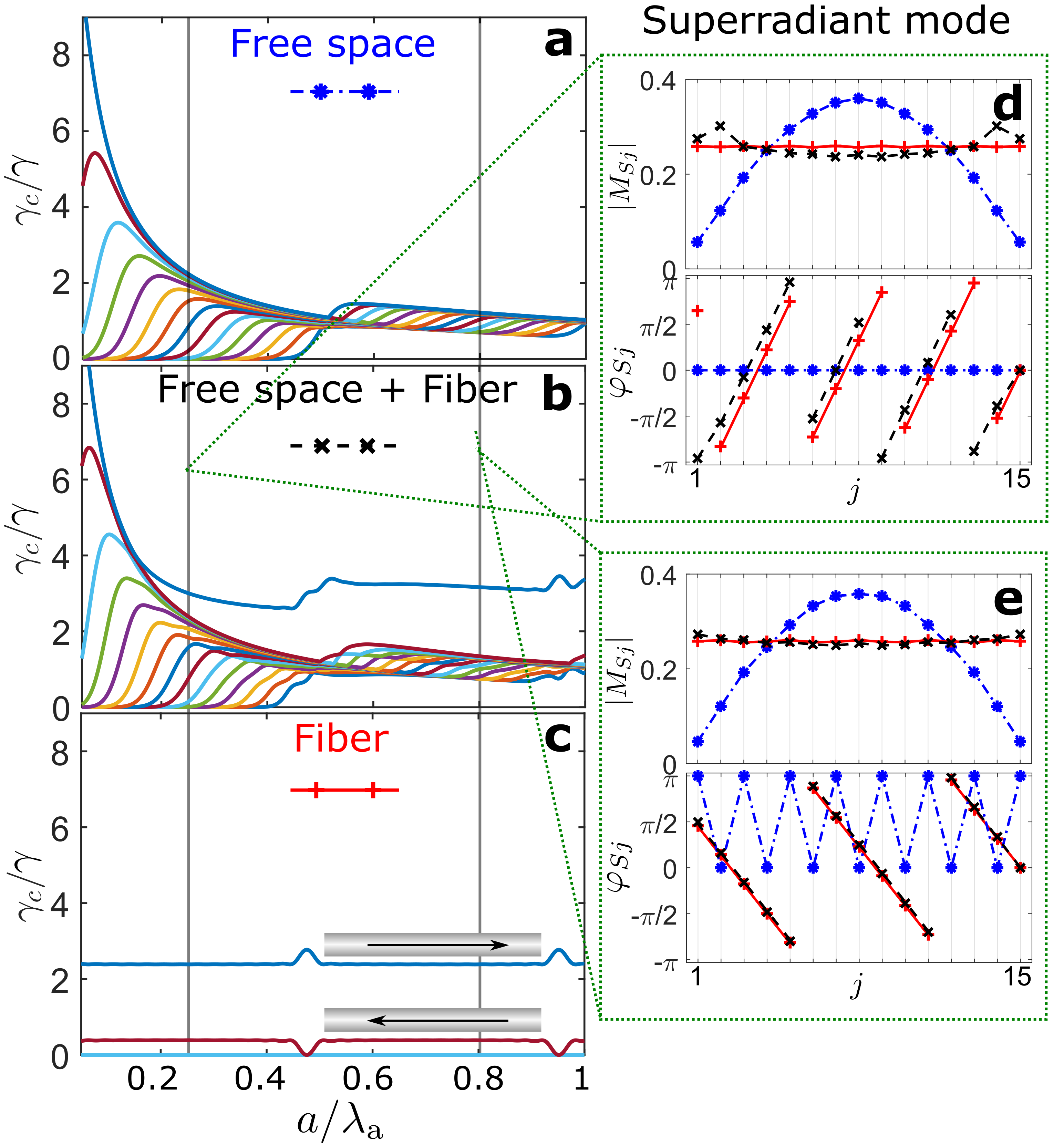}
\caption{\textbf{Hybridization of decay modes.} Collective decay rates $\gamma_c$ for a chain of $N = 15$ atoms with transition wavelength $\lambda_\mathrm{a} = 1\mu$m \textbf{a:} in free space, \textbf{b:} considering both couplings to unguided free field modes and guided modes in a silica nanofiber at a distance $h = 100$nm and \textbf{c:} only to the guided modes of the fiber, where the rightward and leftward-propagating modes are indicated. The decay rates are scaled by the single atom decay rate $\gamma$. \textbf{d} and \textbf{e:} Magnitude $|M_{Si}|$ and phase $\varphi_{Si}$ of the superradiant decay mode's spatial profile for a chain with $a=250$ and $800$ nm, respectively, in free space (blue stars), coupled to the nanofiber guided modes (red plusses) and to both (black crosses).}\label{fig:decayRatesAndModes_varyA}
\end{figure}

\textit{Collective decay modes.} In order to gain an understanding of the collective decay modes, we first consider the (free space) situation where the fiber is absent (Fig. \ref{fig:decayRatesAndModes_varyA}a). For small interatomic separation, $a/\lambda_\mathrm{a}\ll1$, the off-diagonal elements of $\Gamma_{ij}$ become comparable to the diagonal ones and  superradiant ($\gamma_c\gg\gamma$) and subradiant ($\gamma_c\ll\gamma$) modes emerge. As the distance between the atoms is increased, the magnitude of the off-diagonal elements quickly decays, and hence all collective decay rates approach the single-atom value, $\gamma$.

In Fig. \ref{fig:decayRatesAndModes_varyA}b we show the same collective decay rates for an atomic chain at $h = 100$nm from a silica nanofiber with radius $r_\mathrm{f} = 220$nm. When the spacing between the atoms is much smaller than the transition wavelength, we observe that the collective decay rates do not change significantly from the ones in Fig. \ref{fig:decayRatesAndModes_varyA}a. As $a/\lambda_\mathrm{a}$ is increased, however, a superradiant mode with enhanced decay rate splits from the rest. This superradiant mode corresponds to a guided rightward-propagating decay mode that emerges due to the presence of the fiber. This mode can also be observed in Fig. \ref{fig:decayRatesAndModes_varyA}c, where the decay rates in the absence of free field are depicted. Note, that here we also identify a second (leftward-propagating) mode, which we will discuss later.

In order to gain an understanding of the nature of the superradiant mode, we show the magnitude and phase of its spatial profile $M_{Sj}= |M_{Sj}| \mathrm{e}^{i \varphi_{Sj}}$ in Figs. \ref{fig:decayRatesAndModes_varyA}d and e for two values of $a/\lambda_\mathrm{a}$. The profile of the superradiant eigenmode is given for the three cases depicted in Figs. \ref{fig:decayRatesAndModes_varyA}a, b and c. For very small interatomic separation the spatial profile of all decay modes is independent of the presence of the fiber. As $a/\lambda_\mathrm{a}$ increases, the collective decay rates corresponding to the two most superradiant modes cross, and the profile of the "hybrid" superradiant mode becomes similar to the fully guided one (see black crosses and red plusses in panel d). For larger distances between the atoms (panel e), this hybridized superradiant mode is almost completely formed by one of the fiber guided modes, as the virtually identical mode profiles in Fig. \ref{fig:decayRatesAndModes_varyA}e show.

Let us further analyze the phase profile of this superradiant mode, which will be key for understanding the collective enhancement of the rate and chirality of the guided photon emission. For the parameters chosen here, 72\% of the photons that are emitted into the nanofiber from each single atom propagate rightwards, i.e., the single-atom guided coupling is chiral (see Appendix B). This symmetry breaking in the emission is manifested in the superradiant mode as a phase gradient across the chain. The phase difference between nearest neighbor atoms is $\Delta\phi=a\beta_\mathrm{f}$, with $\beta_\mathrm{f}$ being the propagation constant of the light inside the nanofiber. The value of $\beta_\mathrm{f}$ is close to $2\pi/\lambda_\mathrm{a}$. Thus, every time that $a$ crosses a multiple of $\lambda_\mathrm{a}/2$ an apparent change of sign of the phase gradient takes place (see, e.g. panels d and e in Fig. \ref{fig:decayRatesAndModes_varyA}). In order to account for this aliasing, we rewrite the phase difference for nearest neighbors as $\Delta\phi=a\beta_\mathrm{f}-2\pi n$, with $n=0,1,\dots$ being the integer part of $a/(\lambda_\mathrm{a}/2)$. Note as well that for the leftward-propagating mode (second highest decay rate state in Fig. \ref{fig:decayRatesAndModes_varyA}c), the phase gradient has the opposite sign.

\begin{figure*}[ht]
\includegraphics[width=0.9\textwidth]{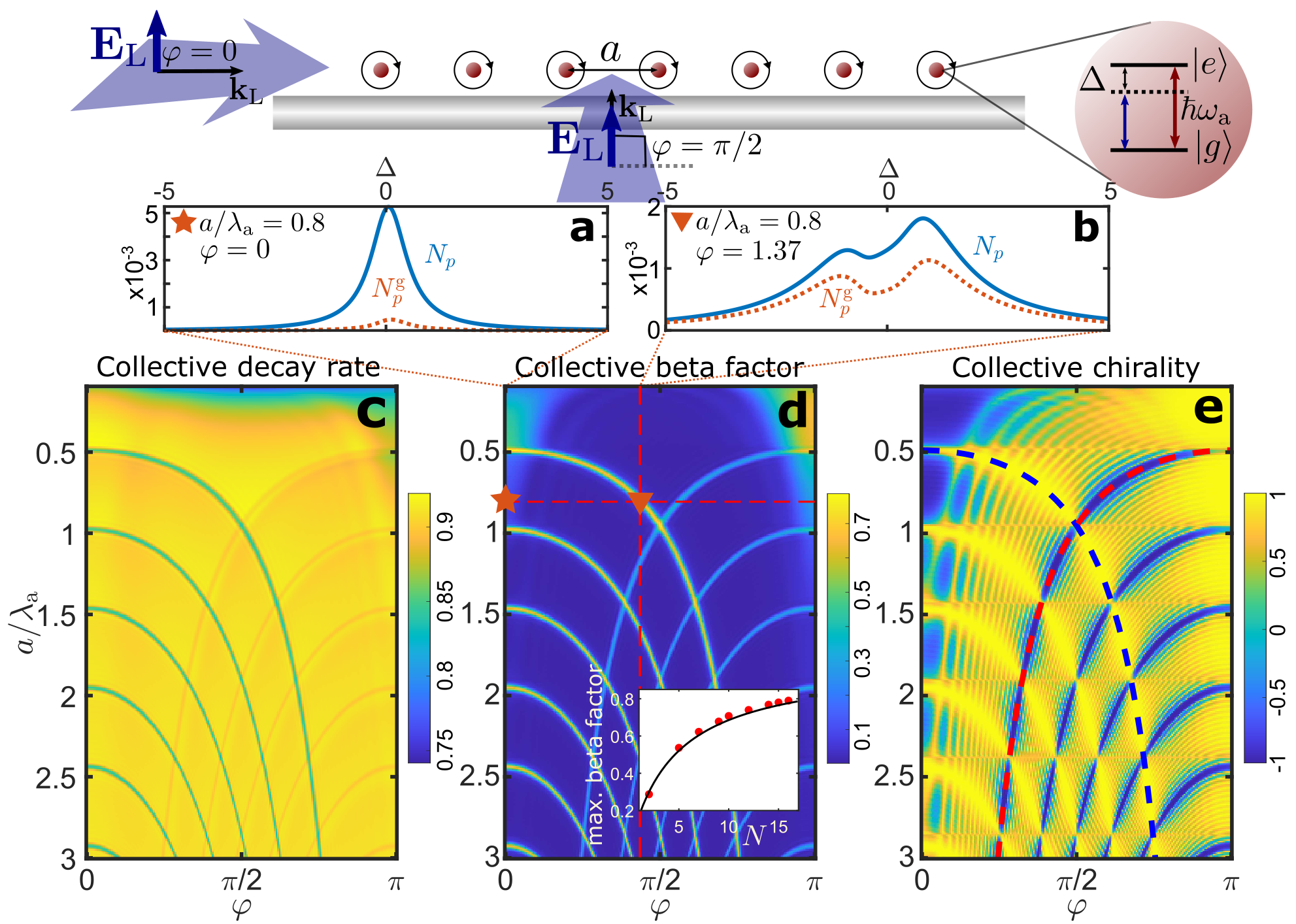}
\caption{\textbf{Collectively enhanced emission properties.} Scheme for the laser excitation of a chain of atoms located at $h=100$nm from a fiber with radius $r_\mathrm{f}=220$nm. The Rabi frequency of the external laser field is $\Omega_\mathrm{L}=\gamma/100$, and the angle between the k-vector of the laser and the orientation of the chain is $\varphi$. \textbf{a} and \textbf{b:} Total photon emission rate $N_p$ (blue solid line) and guided photon emission rate $N_p^\mathrm{g}$ (red dotted line) as a function of $\Delta$ (in units of the single atom decay rate $\gamma$) for $N=15$ at the two points shown in panel \textbf{d}. \textbf{c:} Collective decay rate. \textbf{d:} Collective beta factor. Inset shows the maximum collective beta factor as a function of the system size $N$, red dots. The black line shows  $N\Gamma^\mathrm{g}/(N\Gamma^\mathrm{g}+\Gamma^\mathrm{u})$ (see text).  \textbf{e:} Collective chirality. Shown for comparison is the expression (\ref{eq:fit}) for $n=1$ (dashed lines).} \label{fig:scattering}
\end{figure*}

\textit{Collectively enhanced beta factor.} The excitation of the superradiant mode which we just analyzed gives rise to an enhancement of both the rate and chirality of photon emission into the nanofiber. In order to investigate how this can be tested experimentally, we consider the response of the atomic chain when driven by a weak laser field, $\mathbf{E}_\mathrm{L}(\mathbf{r})=E_\mathrm{L}e^{i\mathbf{k}_\mathrm{L}\cdot\mathbf{r}}\hat{\mathbf{\varepsilon}}_\mathrm{L}$, with polarization $\hat{\mathbf{\varepsilon}}_\mathrm{L}$ in the $x$-direction, and detuned from the $\left|g\right>\to\left|e\right>$ transition by $\Delta$ (see scheme in Fig. \ref{fig:scattering}) \footnote{Note that here for the two-level approximation to be valid in a realistic multilevel atom, one can consider the presence of a uniform magnetic field in the $y$-direction that splits the levels such that the laser is only resonant with the desired transition}. We imprint a phase pattern with the laser by tuning the angle $\varphi$ between the laser momentum $\mathbf{k}_\mathrm{L}$, and the chain. In particular, in order to match the phase profile of the rightward (superradiant), and leftward-propagating states that we introduced in the previous section, the laser angle must satisfy
\begin{equation}\label{eq:fit}
\cos{\varphi}=\pm\left(\frac{n\lambda_\mathrm{a}}{a}-\frac{\lambda_\mathrm{a}}{\lambda_\mathrm{f}}\right),
\end{equation}
with $\lambda_\mathrm{f}=2\pi/\beta_\mathrm{f}$ and $n=1,2,\dots$.

For the subsequent analysis, we define the total photon emission rate as \cite{lekien2008} $N_p(\Delta)=\sum_{ij}\Gamma_{ij} \left<\sigma_i^\dag\sigma_j\right>_\mathrm{ss}$, where $\left<\dots\right>_\mathrm{ss}$ denotes the expectation value in the stationary state. This expectation value can be easily found in the limit of weak laser driving within the single-excitation subspace, as it is described in Appendix C. Analogously, the photon emission rate into the guided modes is given by $N_p^\mathrm{g}(\Delta)=\sum_{ij}\Gamma_{ij}^\mathrm{g} \left<\sigma_i^\dag\sigma_j\right>_\mathrm{ss}$, where the coefficients $\Gamma_{ij}^\mathrm{g}$ contain the couplings into the guided modes only. Moreover, we define the collective decay rate as the integral over the detuning of $N_p(\Delta)$. Similarly, we define the collective beta factor as the ratio between the total photon emission rate into the nanofiber (again integrated over $\Delta$) and the collective decay rate. Finally, the collective chirality is obtained by breaking down the emission rates into the right and left directions of the nanofiber. We define it as the difference between the guided photon emission rate into the rightward and leftward-propagating modes divided by the total photon emission rate into the nanofiber (see Appendix D).

In Figs. \ref{fig:scattering}a and b, we show $N_p$ and $N_p^\mathrm{g}$ for fixed values of the Rabi frequency of the laser $\Omega_\mathrm{L}\propto\left|E_\mathrm{L}\right|^2$ and the ratio $a/\lambda_\mathrm{a}$ for a chain of $N=15$ atoms. The observed behavior strongly depends on the laser angle $\varphi$. For $\varphi=0$ (panel a), both the total and the guided photon rates have a characteristic Lorentzian shape only slightly shifted away from $\Delta=0$. Most importantly, the fraction of emission into the guided modes here is small for all values of the detuning. However, at $\varphi=1.37$ [solution of the equation \eqref{eq:fit} with $n=1$] this fraction is enhanced considerably, which is due to the angle of the laser momentum matching the phase profile of the superradiant state shown in Fig. \ref{fig:decayRatesAndModes_varyA}e. Note, that while the superradiant mode is an eigenstate of the coefficient matrix $\Gamma_{ij}$, it is not an eigenstate of the effective Hamiltonian $H_\mathrm{eff}=\sum_{i\neq j}\left(V_{ij}-i\Gamma_{ij}/2\right)$ that describes the dynamics of the system (see Appendix C), which leads to the splitting of the superradiant peak into two, shifted away from resonance with different signs of the detuning.

In Figs. \ref{fig:scattering}c-e we show the collective decay rate, beta factor, and chirality, as a function of $a/\lambda_\mathrm{a}$ and the laser angle $\varphi$. One clearly observes a collective modification of all quantities when the mode matching condition \eqref{eq:fit} is met, visible in a marked pattern of lines. First, the collective decay rate becomes smaller along these lines. Second, as predicted above, the collective beta factor is increased dramatically, particularly when the laser matches the most superradiant mode. This enhancement becomes more pronounced as the system size is increased, growing approximately as $N\Gamma^\mathrm{g}/(N\Gamma^\mathrm{g}+\Gamma^\mathrm{u})$ \cite{lekien2008}, with $\Gamma^\mathrm{g/u}$ being the single atom emission rates into the guided and unguided modes (see Appendix B). This can be observed in the inset of Fig. \ref{fig:scattering}d, where the maximum collective beta factor is shown as a function of $N$. Finally, the enhancement of the beta factor is accompanied by a dramatic enhancement of the chirality of the emission from its single-atom value ($0.72$ in the example shown here) to $0.999$. I.e., the emission becomes virtually unidirectional. Note that the laser angle can be chosen such that the second guided leftward-propagating mode is excited (with a smaller beta factor), achieving negative collective chirality ($-0.999$ in Fig. \ref{fig:scattering}e). Finally, we find that when $a$ is an integer multiple of $\lambda_\mathrm{f}$, the chirality recovers its single-atom value and the collective beta factor reaches its maximum at $\varphi=\pi/2$. The reason is that here the phase profile of the two guided modes becomes flat ($\Delta\phi=\pm2\pi n$) and, thus, the laser matches both modes simultaneously when its momentum is perpendicular to the chain.

\textit{Conclusion and outlook.} We show that the emission from a chain of atoms into the guided modes of a nanofiber can be collectively enhanced and made perfectly unidirectional. This can be achieved by mode matching the phase profile of an excitation laser to the phase gradient of a superradiant state emerging from the atom-fiber coupling. The parameters used in this work are achievable in current experimental setups \cite{mitsch2014,scheucher2016,solano2017,dareau2018}.

A natural continuation to this work will be to investigate the properties and photon counting statistics (e.g. two-time correlations) of the light emitted into the nanofiber. Moreover, the challenge is to go beyond the weak excitation limit and to understand the fate of the collectively enhanced photon emission when the atoms are driven closer to saturation.

\begin{acknowledgments}
The authors acknowledge fruitful discussions with Philipp Schneewei\ss \hspace{0.1mm} and all members of the ErBeStA consortium. The research leading to these results has received funding from the European Union's H2020 research and innovation programme [Grant Agreement No. 800942 (ErBeStA)] and EPSRC [Grant No. EP/M014398/1]. IL acknowledges support from the "Wissenschaftler-R\"{u}ckkehrprogramm GSO/CZS“ of the Carl-Zeiss-Stiftung and the German Scholars Organization e.V.. BO was supported by the Royal Society and EPSRC [Grant No. DH130145]. BL is supported by the Leverhulme Trust through the research project Grant No. RPG-2018-213.
\end{acknowledgments}

\appendix

\section{Appendix A: Master equation terms calculation}

The interaction coefficients can be calculated as
\begin{equation}\label{eq:Vab}
V_{ij} = - \mathcal{P} \sum_{\nu} \left[ \frac{G^{}_{\nu i} G^*_{\nu j}}{\omega - \omega_\mathrm{a}} + (-1)^{\delta_{ij}} \frac{\tilde{G}^*_{\nu i} \tilde{G}^{}_{\nu j}}{\omega + \omega_\mathrm{a}} \right],
\end{equation}
and
\begin{equation}\label{eq:Gab}
\Gamma_{ij} = 2 \pi \sum_{\nu} G^{}_{\nu i} G^*_{\nu j} \delta{(\omega - \omega_a)}.
\end{equation}
Here, $\mathcal{P}$ denotes the Cauchy principal value, $\delta_{ij}$ denotes the Kronecker delta function, and $\sum_{\nu} = \sum_\mathrm{g} + \sum_\mathrm{u}$ is a generalised sum over guided and unguided modes. The sum over the guided modes reads $\sum_\mathrm{g} = \int^{\infty}_0 d\omega \sum_{fl}$, where $\omega$ is the mode frequency, and $l = \pm1$ denotes counterclockwise or clockwise polarization. Finally, $f = \pm1$ denotes whether the mode propagates in the $+z$ or $-z$ direction, such that the sum over the guided modes can be broken down into the two directions along the fiber. For the unguided modes, we have $\sum_\mathrm{u} = \int^\infty_0 d \omega \int^{k}_{-k} d \beta \sum_{ml}$, where $\beta$ is a continuous variable in the range $-k < \beta < k$ with $k = \omega_\mathrm{a} / c$, $m = 0, \pm 1, \pm 2, ...$ denotes the mode order, and $l = \pm1$ again denotes the mode polarization. $G_{\nu i}$ are the coupling strengths between the atom $i$ and the mode $\nu$ which, for the guided and unguided modes, respectively, are given by
\begin{align} \label{eq:Gmi}
&G_{\mathrm{g}i} = \sqrt{\frac{\omega \beta_\mathrm{f}^{'}}{4 \pi \hbar \varepsilon_0}} \left[ \mathbf{d} \cdot \mathbf{e}^{(\mathrm{g})}(r_i, \phi_i) \right] \mathrm{e}^{i(f \beta_\mathrm{f} z_i + l \phi_i)}, \nonumber \\
&G_{\mathrm{u}i} = \sqrt{\frac{\omega}{4 \pi \hbar \varepsilon_0}} \left[ \mathbf{d} \cdot \mathbf{e}^{(\mathrm{u})}(r_i, \phi_i) \right] \mathrm{e}^{i(\beta z_i + m \phi_i)}. 
\end{align}
Here, $\mathbf{e}^{(\nu)}$ denotes the profile function of the electric field part of $\nu$, the explicit forms of which can be found, e.g., in \cite{lekien2017}. Since we consider a single fundamental guided mode, $\mathrm{HE}_{11}$, the value of the longitudinal propagation constant of the mode $\beta_\mathrm{f}$ and its derivative $\beta_\mathrm{f}' = \mathrm{d}\beta_\mathrm{f}/\mathrm{d}\omega$, which must be determined numerically as the solution of an eigenvalue equation \cite{snyder1983}, are the only ones in the range $k < \beta_\mathrm{f} \leq k n_\mathrm{f}$. The tilde in this notation (e.g. $\tilde{G}^{}_{\nu j}$) serves to indicate that the dipole moment $\mathbf{d}$ is to be replaced with its complex conjugate. Note that the modes $\mathbf{e}^{(g)}$ and $\mathbf{e}^{(u)}$ are normalized in different manners, leading to dimensions of inverse distance ($[m^{-1}]$) and root-time inverse distance ($[t^{1/2} m^{-1}]$), respectively \cite{lekien2005}. As required, the generalized sums have the same dimensions for both.

\section{Appendix B: Modified single-atom emission}\label{section:singleAtomDecay}

We characterize the emission from a single atom near a nanofiber via three quantities: the total decay rate ($\Gamma$), the fraction of the total photon emission that enters into the guided modes of the nanofiber (beta factor), and the degree of chirality of the emission ($C$).

The total decay rate $\Gamma$ is obtained from \eqref{eq:Gab} as $\Gamma_{ii}\equiv\Gamma$ (note that in the chain all atoms sit at the same height $h$ above the nanofiber). This rate is the sum of the rates into the guided and unguided modes
\begin{equation*}
\Gamma^{\mathrm{g}/\mathrm{u}} = 2 \pi \sum_{\mathrm{g}/\mathrm{u}} G_{(\mathrm{g}/\mathrm{u})} G^*_{(\mathrm{g}/\mathrm{u})} \delta{(\omega - \omega_\mathrm{a})}.
\end{equation*}
We will thus define the beta factor as $\Gamma^\mathrm{g}/\Gamma$. To quantify the degree of directionality or \textit{chirality} of the decay into the nanofiber, we define
\begin{equation}
C=\frac{\Gamma^\mathrm{g}_\mathrm{R} - \Gamma^\mathrm{g}_\mathrm{L}}{\Gamma^\mathrm{g}},
\end{equation}
with R and L representing the rightward and leftward-propagation direction of the mode, respectively. Here, the rate into the guide modes has been broken down as $\Gamma^\mathrm{g} = \Gamma^\mathrm{g}_\mathrm{R} + \Gamma^\mathrm{g}_\mathrm{L}$ with
\begin{equation}
\Gamma^\mathrm{g}_\mathrm{R/L} = \sum_{l} \frac{\omega_\mathrm{a} \beta^{'}|_{\omega_\mathrm{a}}}{2 \hbar \varepsilon_0} |\mathbf{d} \cdot \mathbf{e}^{(\omega_\mathrm{a}, \pm1, l)}|^2.
\end{equation}
Note that if the transition dipole moment is purely real (for example in the case of a linearly polarized dipole), then the coupling into each of the two guided modes (that possess circular polarization) propagating with opposite directions in the nanofiber is the same and there is no chirality ($C=0$). For this reason, we will consider only circularly polarized dipole moments.

\begin{figure}
\includegraphics[width=\columnwidth]{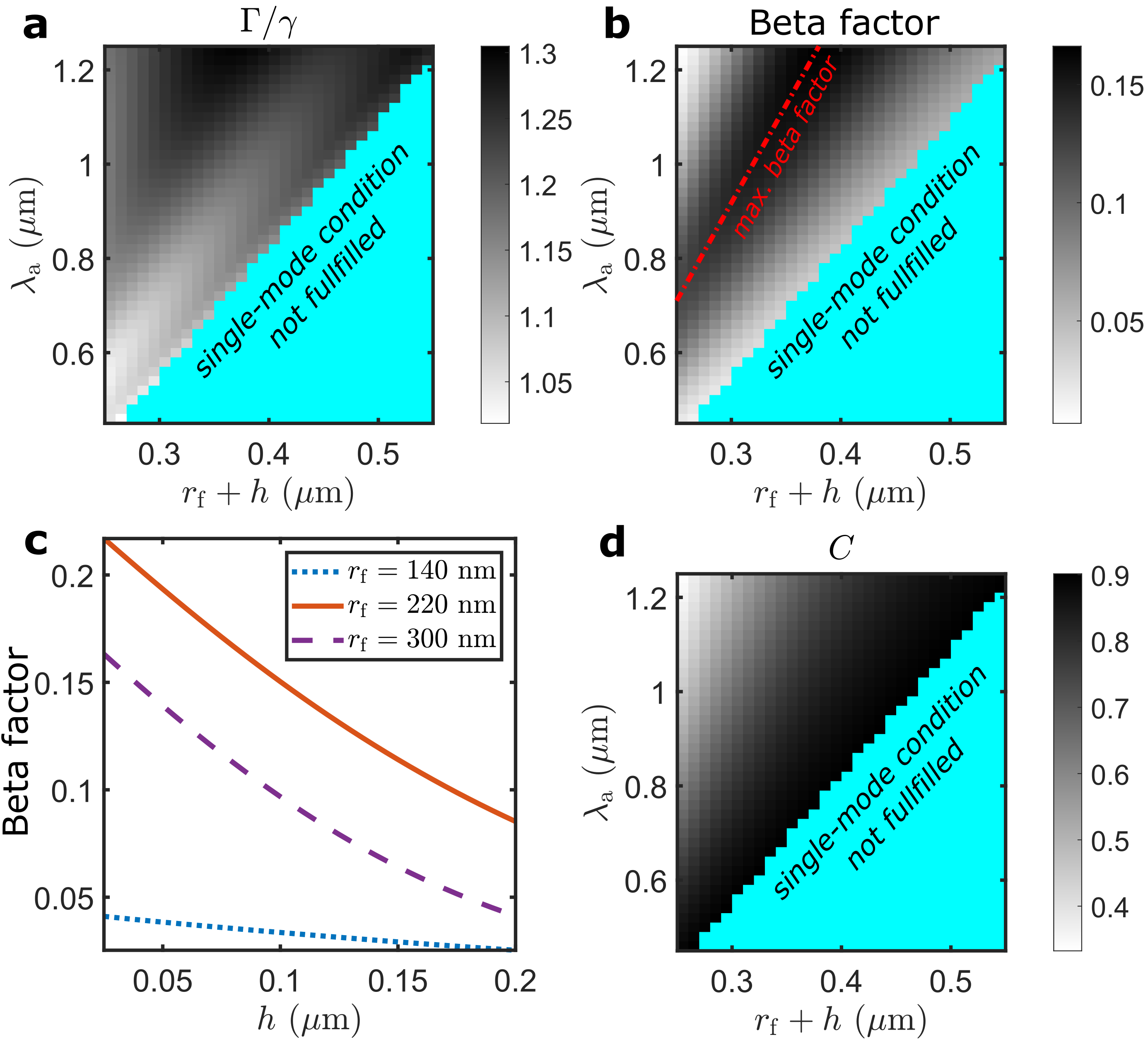}
\caption{\textbf{Modified single atom decay properties.} \textbf{a:} Single-atom decay rate $\Gamma$ for a circularly polarised emitter placed $h = 100$nm from a silica nanofiber. The decay rate is scaled by the one in vacuum, $\gamma$, and plotted varying the transition wavelength $\lambda_\mathrm{a}$ and the fiber radius $r_\mathrm{f}$ (in the blue shaded area the single-mode condition is not fulfilled). \textbf{b:} Beta factor. The red line indicates the value of the nanofiber radius where the strongest coupling to the guided modes is found for each wavelength. \textbf{c:} Beta factor for various fixed values of $r_\mathrm{f}$ as a function of $h$ at $\lambda_\mathrm{a}=1\mu$m. \textbf{d:} Chirality of the decay into the guided modes.}\label{fig:singleAtomDecay_CircDip}
\end{figure}

The overall modification of the decay rate, $\Gamma/\gamma$, is shown in Fig. \ref{fig:singleAtomDecay_CircDip}a for a single atom with circularly polarized transition dipole moment $\mathbf{d}= (d / \sqrt{2})(i,0,-1)$ (sketched in Fig. \ref{fig:System}) placed at a distance $h = 100$nm above a silica nanofiber. The largest value of the varying nanofiber radius $r_\mathrm{f}$ for each wavelength is determined by the single-mode condition. The exact dependence of $\Gamma/\gamma$ with the nanofiber radius and wavelength is not straightforward. However, we do observe that in all cases decay rate is increased with respect to the free space value by the presence of the nanofiber, and that in general this effect is strongest for long transition wavelengths (i.e., when the atom-fiber separation $h$ is much smaller than the transition wavelength).

The beta factor (Fig. \ref{fig:singleAtomDecay_CircDip}b), displays a clearer trend: it has a constant value along the lines $(r_\mathrm{f}+h)/\lambda_\mathrm{a}=\mathrm{constant}$, achieving a maximum approximately at $(r_\mathrm{f}+h)/\lambda_\mathrm{a}\approx1/4$ (red dashed-dotted line). Overall, we also note that the coupling to the guided modes is relatively weak, not increasing above 20\% over the plotted parameter range. Even reducing the distance to the fiber, $h$, to values as small as a few tens of nanometers has only a limited effect on the maximum value achievable for the beta factor, as is reflected in Fig. \ref{fig:singleAtomDecay_CircDip}c.

We investigate the directionality of the emission into the guided modes, $C$, in Fig. \ref{fig:singleAtomDecay_CircDip}d. One can observe that the chirality of the emission is approximately proportional the ratio between the nanofiber radius and the wavelength $r_\mathrm{f}/\lambda_\mathrm{a}$, achieving values up to 0.9 close to the limit of the validity of the single-mode condition. Note, moreover, that in the example shown the chirality is positive, meaning that the guided emission goes prominently into the rightward-propagating guided mode.

\begin{figure}
\includegraphics[width=\columnwidth]{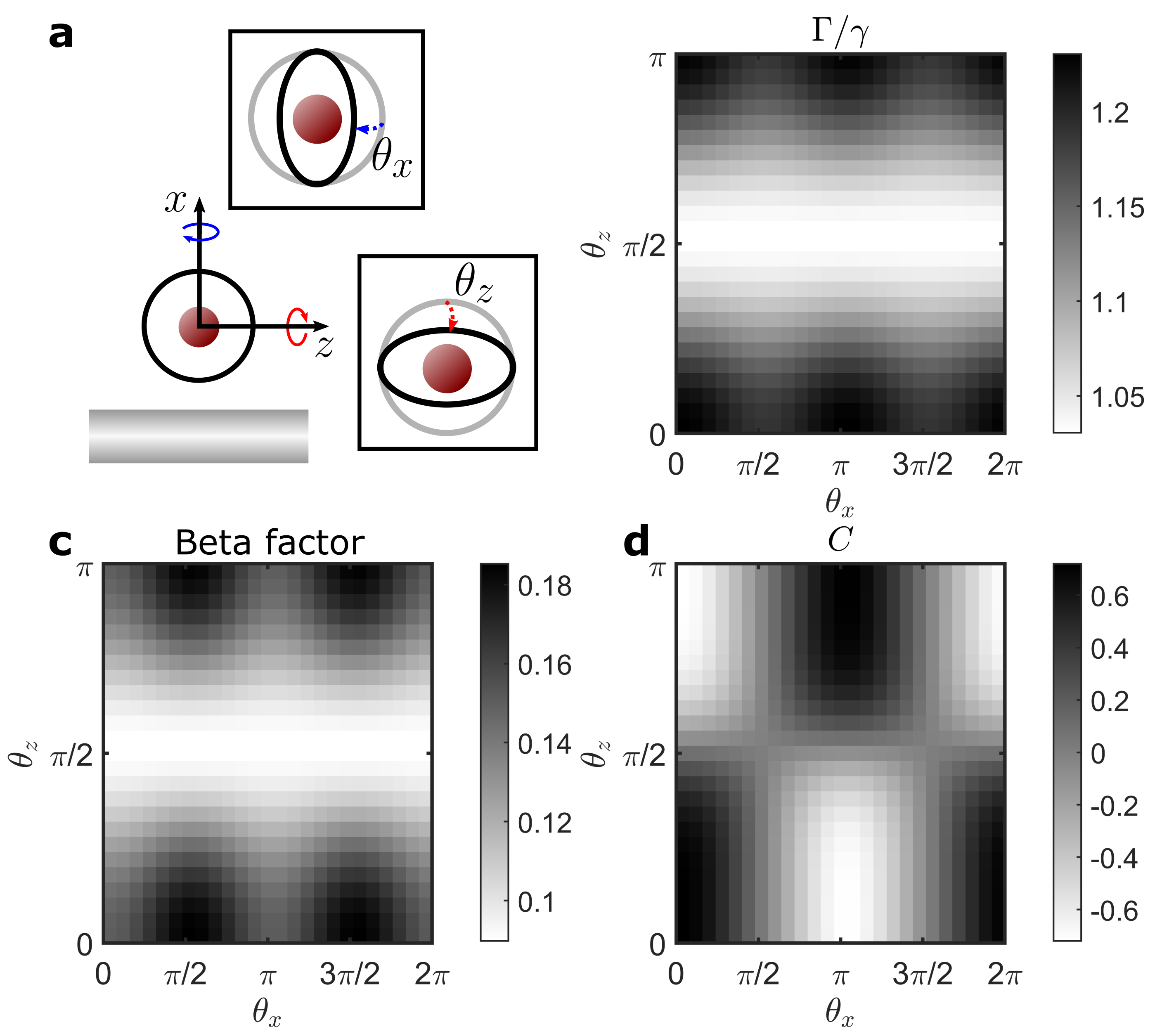}
\caption{\textbf{Role of the polarization direction.} \textbf{a:} Circularly polarised emitter with transition wavelength $\lambda_\mathrm{a} = 1\mu$m placed $h = 100$nm above a silica nanofiber with radius $r_\mathrm{f}= 0.22\lambda_\mathrm{a}$. The transition dipole moment is rotated about the axis parallel to the fiber by an angle $\theta_z$, or about the axis parallel to the atom-fiber separation by an angle $\theta_x$. \textbf{b:} Modified decay rate. \textbf{c:} Beta factor. \textbf{d:} Chirality of the decay into the guided modes.}\label{fig:singleAtomDecay_DipRot}
\end{figure}

Up to now, we have considered a fixed direction of the circularly polarized transition dipole moment $\mathbf{d}$, contained in the $xz$-plane. Different directions of this dipole moment modify dramatically the coupling of the atom both to the free radiation and the guided modes. This effect is explored in Fig. \ref{fig:singleAtomDecay_DipRot} for an atom with transition wavelength $\lambda_\mathrm{a} = 1\mu$m, where the nanofiber radius $r_\mathrm{f} = 0.22\lambda_\mathrm{a}$ is chosen such that the coupling to the guided modes is strongest for $h=100$nm (red line in Fig. \ref{fig:singleAtomDecay_CircDip}b). The dipole moment can be rotated about the axis parallel to the nanofiber by an angle $\theta_z$, and about the axis parallel to the atom-fiber separation by an angle $\theta_x$.

We see that rotating away from $\theta_z  = 0$ leads to a reduction in both the modified decay rate $\Gamma$, which gets closer to $\gamma$, and the beta factor. The beta factor increases to a maximum at $\theta_x=\pi/2$, when the orientation of the dipole moment matches that of the guided mode. The overall decay rate is reduced under this same rotation, as the dipole couples less to the $r$ component of the electric field and more to the weaker $\phi$ component. The chirality is also reduced, falling to zero at $\theta_x = \pi/2$ (see Fig. \ref{fig:singleAtomDecay_DipRot}d) as the dipole moment is then aligned perpendicular to the mode propagation direction, leading to a left/right symmetry in the decay.

\section{Appendix C: Stationary state in the single-excitation subspace}

The wave function that describes the state of the system can be written in the low excitation limit as $\left|\psi(t)\right>=c_G(t)\left|G\right>+\sum c_e^i(t)\left|e\right>_i$, where $\left|G\right>\equiv\left|g\right>_1\otimes\left|g\right>_2\dots\otimes\left|g\right>_N$ and $\left|e\right>_i\equiv\left|g\right>_1\otimes\left|g\right>_2\dots\left|e\right>_i\dots\otimes\left|g\right>_N$ are the many-body ground and single excitation states, respectively. In this subspace, the photon emission rates depend only on the stationary state value of the coefficients $c_e^i$, which can be found as a solution of the equation
\begin{equation}
\left(\Delta+i\frac{\Gamma}{2}\right) \mathbf{c}_e^\mathrm{ss}=H_\mathrm{eff}\mathbf{c}_e^\mathrm{ss}+\Omega_\mathrm{L}\mathbf{v}.
\end{equation}
Here, $\mathbf{c}_e^\mathrm{ss}$ is the vector that contains the stationary state coefficients $c_e^i$, $H_\mathrm{eff}=\sum_{i\neq j}\left(V_{ij}-i\Gamma_{ij}/2\right)$, $\Omega_\mathrm{L}$ is the Rabi frequency of the laser, and $\mathbf{v}=\sum_{i}e^{i\mathbf{k}_\mathrm{L}\cdot\mathbf{r}_i}$.

\section{Appendix D: Collective emission characterization}

The characterization of the emission from the weakly driven atomic array is done through three quantities: The collective decay rate, the collective beta factor and the collective chirality. Here we introduce the definitions that are only indicated in the main text.

The collective decay rate $\Gamma_\mathrm{C}$ is defined as
\begin{equation}
\Gamma_\mathrm{C}=\int_{-\infty}^\infty d\Delta N_p(\Delta).
\end{equation}
In order to define a collective collective beta factor, we define first the collective guided rate as
\begin{equation}
\Gamma^\mathrm{g}_\mathrm{C}=\int_{-\infty}^\infty d\Delta N^\mathrm{g}_p(\Delta),
\end{equation}
such that the collective beta factor is defined as the fraction of the photons that are emitted into the nanofiber, i.e. the $\Gamma^\mathrm{g}_\mathrm{C}/\Gamma_\mathrm{C}$. Finally, the collective chirality is similarly obtained by breaking down the emission rates into the right and left directions of the nanofiber, i.e. $N^\mathrm{g}_p(\Delta)=N^\mathrm{gR}_p(\Delta)+N^\mathrm{gL}_p(\Delta)$. The collective emission rates into each direction are thus given by
\begin{equation}
\Gamma^\mathrm{gR/L}_\mathrm{C}=\int_{-\infty}^\infty d\Delta N^\mathrm{gR/L}_p(\Delta),
\end{equation}
and the collective chirality is defined as
\begin{equation}
C_\mathrm{C}=\frac{\Gamma^\mathrm{gR}_\mathrm{C}-\Gamma^\mathrm{gL}_\mathrm{C}}{\Gamma^\mathrm{gR}_\mathrm{C}+\Gamma^\mathrm{gL}_\mathrm{C}}.
\end{equation}
Note that all three quantities defined here reduce to their single-atom counterparts in the limit of $N=1$ (Appendix B).


\end{document}